\documentclass{rspublic}

\usepackage{graphicx}
\newcommand{\lag}{\langle}
\newcommand{\rag}{\rangle}
\newcommand{\Map}{M_{\rm ap}}

\begin{document}
\title[Weak Lensing Surveys]{Cosmology with Weak Lensing Surveys}
\author[D.Munshi and P.Valageas]{Dipak Munshi$^{1,2}$ and Patrick Valageas$^{3}$}
\affiliation{$^{1}$Institute of Astronomy, Madingley Road, Cambridge, CB3 OHA, UK \\
$^{2}$Astrophysics Group, Cavendish Laboratory, Madingley Road, 
Cambridge CB3 OHE, UK \\
$^{3}$ Service de Physique Th\'eorique, 
CEA Saclay, 91191 Gif-sur-Yvette, France} 
\label{firstpage}
\maketitle

\begin{abstract}{Weak lensing - Cosmology: methods - Statistical}
Weak gravitational lensing is responsible 
for the shearing and magnification of the images of high-redshift sources due 
to the presence of intervening mass. Since the lensing effects arise from
deflections of the light rays due to fluctuations of the gravitational
potential, they can be directly related to the underlying density field of
the large-scale structures. Weak gravitational surveys are complimentary
to both galaxy surveys and cosmic microwave background observations
as they probe unbiased non-linear matter power spectra at medium redshift.
Ongoing CMB experiments such as WMAP and future 
Planck satellite mission will measure
the standard cosmological parameters with unprecedented accuracy. The focus 
of attention will then shift to understanding the nature of dark matter 
and vacuum energy: several recent studies suggest that 
lensing is the best method for constraining the dark energy equation of state.
During the next 5 year period ongoing and future weak 
lensing surveys such as the Joint Dark Energy Mission (JDEM, e.g. SNAP) or the
Large-aperture Synoptic Survey Telescope (LSST) will play a major role in
advancing our understanding of the universe in this direction.
In this review article we describe various aspects of probing the matter power 
spectrum and the bispectrum and other related statistics with weak lensing 
surveys. This can be used to probe the background dynamics of the universe 
as well as the nature of dark matter and dark energy. 
\end{abstract}

\section{Introduction}

Gravitational lensing refers to the deflection of light from distant sources by
the gravitational force arising from massive bodies present along the line of 
sight. Such an effect was already raised by Newton in 1704. Indeed,
as derived in textbooks on Newtonian mechanics a particle starting at velocity 
$v$ at large distance from a spherical body of mass $M$ is deflected by an 
angle $\alpha_N=2GM/rv^2$, where $r$ is the impact parameter and we 
assumed that $\alpha_N$ is small. Setting $v=c$ one obtains the ``Newtonian'' 
value for the deflection of light, as calculated by Cavendish around 1784.
As is well known, General Relativity yields twice the Newtonian value, 
$\alpha_{GR}=2\alpha_N$, as obtained by Einstein (1915).
The agreement of this prediction with the deflection of light from distant 
stars by the sun measured during the solar eclipse of 1919 
(Dyson, Eddington \& Davidson 1920) 
was a great success for Einstein's theory and brought General Relativity to 
the general attention (the eclipse allows one to detect stars with a line of 
sight which comes close to the sun).

In a similar fashion, light rays emitted by a distant galaxy are deflected
by the matter distribution along the line of sight toward the observer. This
creates a distortion of the image of this galaxy, which is both sheared and 
amplified (or attenuated). While the gravitational lensing effect due to
a rare massive object like a cluster of galaxies can be very strong and lead to
multiple images, the distortion associated with typical density fluctuations
is rather modest (of the order of $1\%$). Besides, galaxies are not exactly
spherical. Therefore, one needs to average over many galaxies and 
cross-correlate their observed ellipticity in order to extract a meaningfull
signal and this field of investigation is called ``statistical weak 
gravitational lensing''. Thus, the aim of ``weak lensing surveys'' is to obtain
the images of distant galaxies over a whole region on the sky in order to
estimate the coherent shear over large angular scales due to the large scale 
structures of the universe. This allows us to derive information on the
structure and dynamics of the universe 
from the statistical deformation 
of the images of distant galaxies (for more detailed reviews or articles see 
for instance Miralda-Escude 1991, Bartelmann \& Schneider 2001, Mellier 1999, 
Refregier 2003).


In the last few years many studies have managed to detect cosmological shear 
in random patches of the sky (Bacon, Refregier \& Ellis 2000; Bacon et al. 2003; Brown et al. 2003; Hamana et al. 2003; H\"{a}mmerle et al. 2002; Hoekstra, Yee \& Gladders 2002a; Hoekstra et al. 2002a; Jarvis et al.  2002; Kaiser, Wilson \& Luppino 2000; Maoli et al. 2001; Refregier, Rhodes \& Groth 2002; Rhodes, Refregier \& Groth 2001; van Waerbeke et al. 2000; van Waerbeke et al. 2001a; van Waerbeke et al. 2002; Wittman et al. 2000). 
While early studies were primarily concerned with the detection of
a non-zero weak lensing signal, present weak lensing studies are already 
putting constraints on cosmological parameters such as the matter density 
parameter $\Omega_m$ and the normalisation $\sigma_8$ of the power-spectrum
of matter density fluctuations. These works also help to break
parameter degeneracies when used along with other cosmological probes
such as Cosmic Microwave Background (CMBR) observations. In combination with 
galaxy redshift surveys they can be used to study the bias associated with 
various galaxies which will be useful for galaxy formation scenarios thereby
providing much needed clues to the galaxy formation processes. In this article 
we review the recent progress that has been made and various prospects of 
future weak lensing surveys.

\section{Using weak gravitational lensing effects for cosmology}

\subsection{Theory}

Until very recently most of the information about the power spectrum of 
cosmological density perturbations was obtained from large scale galaxy 
surveys and Cosmic Microwave Background Radiation observations. 
However, galaxy surveys only probe the clustering of luminous matter while 
CMBR observations measure the power spectrum at a very early linear stage. 
Weak lensing studies fill in the gap by giving us a direct handle on the 
cosmological power spectrum at a medium redshift in an unbiased way, from 
linear to non-linear scales. As recalled in the introduction, light rays 
emitted by a distant object are deflected because of the inhomogeneites of 
the matter distribution along the line of sight. This implies that a source
located at the angular position ${\bf\theta}$ on the sky is actually 
observed at the angular location ${\bf\theta}+\delta{\bf\theta}$ 
given by the symmetric shear matrix (Schneider, Ehlers \&  Falco 1992; 
Jain, Seljak \& White 2000):
\begin{equation}
\psi_{ij}=\frac{\partial\delta\theta_i}{\partial\theta_j} = 
\left( \begin{array}{cc} \kappa+\gamma_1 & \gamma_2 \\ 
\gamma_{2} & \kappa-\gamma_1 \end{array} \right) = 
2 \int_0^{\chi_s} {\rm d}\chi \frac{{\cal D}(\chi){\cal D}(\chi_s-\chi)}
{{\cal D}(\chi_s)} \partial_i\partial_j \phi(\chi) ,
\label{shearmatrix}
\end{equation}
where we assumed distortions are small. Here $\chi$ is the radial comoving 
distance, ${\cal D}$ is the angular distance and the subscript $s$ refers to
the source redshift, while $\phi$ is the perturbed gravitational potential
(i.e. the source term to Poisson's equation is the perturbed density 
$\rho-{\overline \rho}$ where ${\overline \rho}$ is the mean density of the 
universe). Thus, eq.(\ref{shearmatrix}) clearly shows how the deflection
angles $\delta{\bf\theta}$ are related to the matter distribution along the
line of sight. Of course, in practice we cannot measure the deflections
$\delta{\bf\theta}$ themselves (since usually we do not know the actual 
positions ${\bf\theta}$) but we can measure the distortions $\psi_{ij}$
(assuming we know the shape of the source). Thus, the convergence $\kappa$ 
describes isotropic distortions of images (contractions or dilatations) while 
the shear ${\bf \gamma} = (\gamma_1, \gamma_2)$ describes anisotropic 
distortions. Since galaxies are not spherical we need to average the observed
ellipticities of many galaxies over a given region on the sky to obtain the
coherent weak lensing shear over this area (or its low order moments) 
associated with large scale structures.

\subsection{From galaxy ellipticities to weak lensing shear}

In order to measure the shear $(\gamma_1,\gamma_2)$ from the shape of galaxies
one usually computes the second moments $I_{ij}$ of the galaxy surface
brightness which can be combined to yield its observed ellipticity 
${\bf e}_{\rm obs}$ (Kaiser, Squires \& Broadhurst 1995):
\begin{equation}
I_{ij} = \int {\rm d}^2 \theta \;\; \theta_i \theta_j I({\bf \theta})
W({\bf \theta})  , \hspace{0.5cm}
{\bf e}_{\rm obs} = \left( \frac{I_{11}-I_{22}}{I_{11}+I_{22}} , 
\frac{2 I_{12}}{I_{11}+I_{22}} \right) .
\label{eobs}
\end{equation}
The window function $W({\bf \theta})$ is centered on the galaxy and suppresses
the noise from other parts of the sky. If there were no observational errors
the observed ellipticity ${\bf e}_{\rm obs}$ would be related to the actual
galaxy ellipticity by ${\bf e}$ by: ${\bf e}_{\rm obs}={\bf e}+{\bf \gamma}$
in the linear regime (small lensing distortions). Then, assuming that galaxy
ellipticities are uncorrelated over large distances we can recover the
coherent weak lensing shear ${\bf \gamma}$ by summing over many sources.
Note that in doing so we must also take into account the redshift distribution
of the galaxies. 

In practice, we must first remove the observational noise from 
${\bf e}_{\rm obs}$. Since weak lensing distortions are only of the order of
a few percents the various stages of data reduction need to be performed with 
extreme care. First, it is desirable to have a homogeneous depth over the 
field and as small a seeing as possible. Besides, it is advantageous to have
a wide survey which contains many uncorrelated cells (since we wish to perform
a statistical analysis). Next, after flat fielding (to remove sensitivity 
differences between pixels) one maps the images from ``detector coordinates'' 
to actual astronomical coordinates taking into account telescope distortion
(for instance by using a set of reference stars and modeling the detector 
distortion by a low-order polynomial). Then, one needs to correct for the
asymmetry and smearing of the point-spread function (PSF), which describes 
the response of the imaging system to an object consisting of a perfect point. 
This distortion may be due to atmospheric turbulence, guiding errors and
telescope optics. This can be corrected by measuring the shape of the stars 
present in the field of view, which ought to be spherical and can act as 
point sources for this purpose. Note that in general the PSF can vary with
time and needs to be modeled individually for each image. 
See Kaiser et al. (1999) for a detailed description of image processing.

Elaborate methods have been developed to tackle these issues. The most 
commonly used technique of Kaiser, Squires \& Broadhurst (KSB, 1995) is
based on the measurement of the quadrupole moment of the galaxy surface 
brightness (see eq.(\ref{eobs})) and it treats the PSF convolution analytically
to first order. It was later generalised by several authors 
(Luppino \& Kaiser 1997; Hoekstra et al. 1998). Recent studies have 
thoroughly tested KSB techniques and found them to be sufficient for present 
surveys (e.g. Erben et al. 1991 and Bacon et al. 2001). 
However, a higher level of accuracy will be needed for future generations of 
surveys. Many improvements and alternatives were studied in the recent past 
to deal with these issues by modifying or extending KSB (Kaiser 2000; 
Rhodes, Refregier \& Groth 2000), fitting observed galaxy shapes with 
analytical models (Kuijken 1999) or decomposing galaxy shapes over orthogonal 
basis functions (Bernstein \& Jarvis 2002; Refregier \& Bacon 2003).

Note that weak lensing effects can also be measured through the induced 
magnification associated with the convergence $\kappa$. This changes the 
flux received from a distant source,
its observed size and the number counts of objects.


\subsection{Numerical simulations}

\begin{figure}
\begin{center}
\setlength{\unitlength}{1cm}
\begin{picture}(5,4)(0,0)
\put(-2.25,-.75){\includegraphics{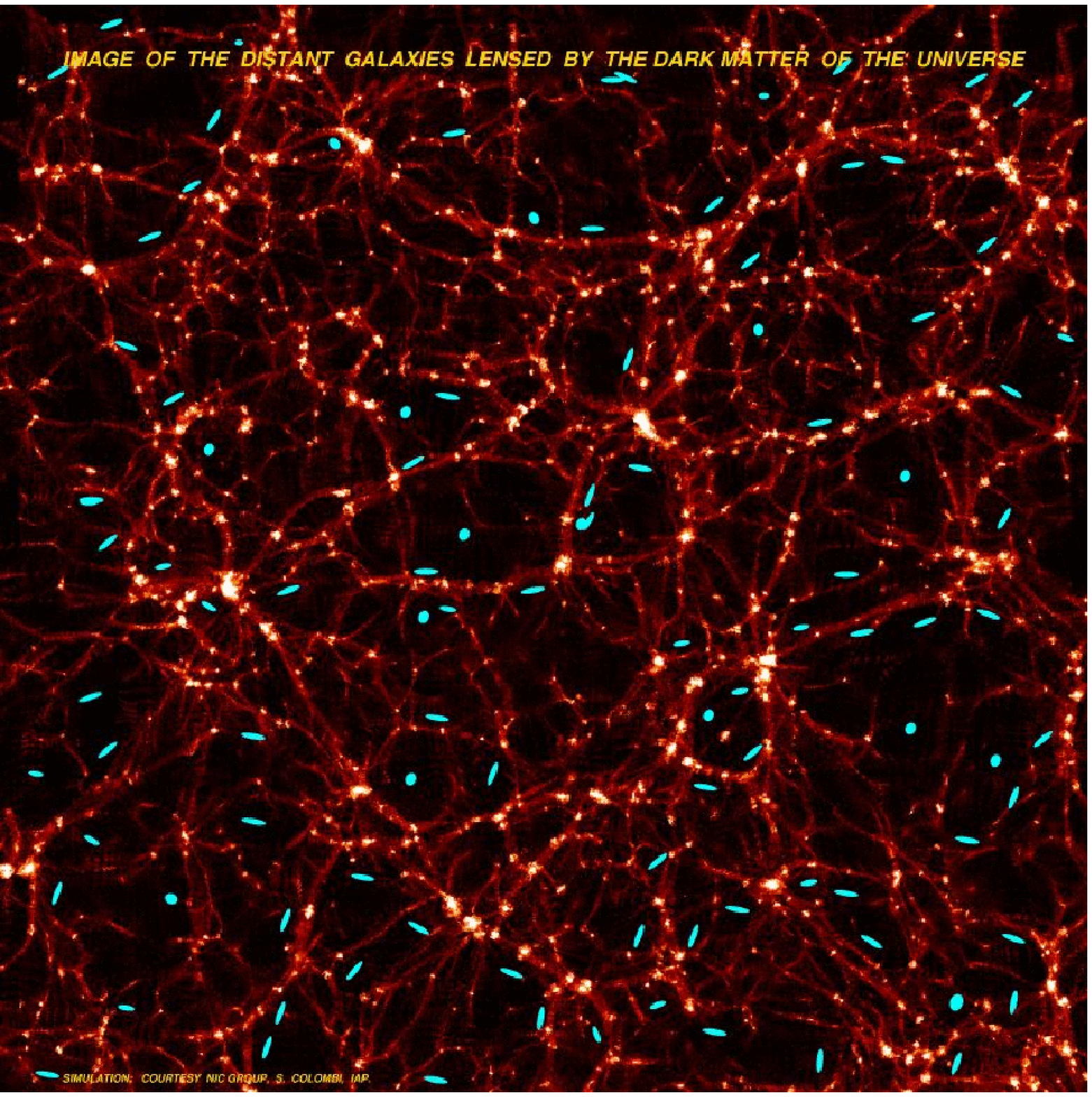}}
\put(2.75,-1.4){\includegraphics{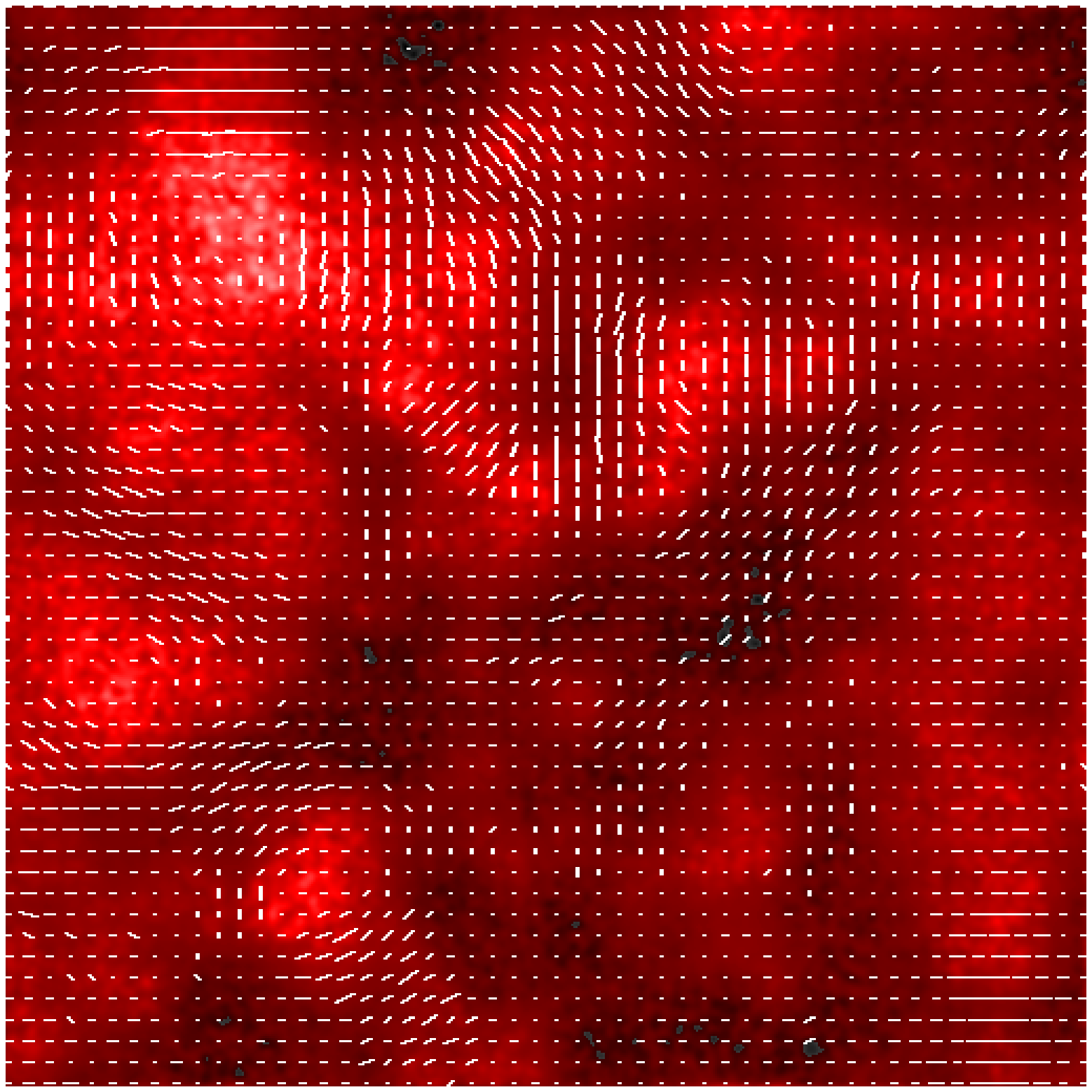}}
\end{picture}
\end{center}
\caption[]{Left panel: The blue elongated disks represent observed 
images of background galaxies. The dark matter filaments from numerical simulations which were used to simulate the survey are also plotted to show 
alignemnets of ellipticites of observed galaxies with underlying
filamentary structures (Figure courtesy:
Stephane Colombi). Right panel: Simulated shear field (short solid lines)
and convergence field (background intensity) are shown (Figure courtesy: 
Patrick Simon) .}
\label{label_fig}
\end{figure}

Once the data analysis has allowed one to obtain a shear map from a weak 
lensing survey, one can use it to put constraints on cosmology by comparing
the observational data with predictions associated with different cosmological
models. 
On large scales where the density contrast is small ($|\delta| \ll 1$ with
$\delta=(\rho-{\overline \rho})/{\overline \rho}$ where ${\overline \rho}$ 
is the mean density of the universe) one can
use an analytical perturbative approach to obtain the properties of weak 
lensing (Bernardeau, van Waerbeke \& Mellier 1997).
However, at small non-linear scales there is no rigorous
analytical method to derive the statistics of the density contrast $\delta$.
Therefore, one needs to use numerical simulations to make the contact between
observations and the relevant cosmological parameters which describe the 
universe. To do so, one first builds a numerical simulation of the growth
of large scale structures in the universe through gravitational instability for
a given set of cosmological parameters. This gives a map of the density 
contrast and of the gravitational potential along a line of sight up to high
redshift (one actually piles up several simulation boxes up to the source
redshift). Then, one can follow the propagation of light through the simulation
boxes (ray tracing technique, Blandford et al. 1991; 
Jain, Seljak \& White 2000; Premadi et al. 2001; 
Wambsganss, Cen \& Ostriker 1998) or integrate the shear matrix along the 
line of sight (Barber et al. 2000; Hamana, Martel \& Futamase 2000).
Finally, by averaging over many lines of sight one can obtain the statistical
properties of weak lensing effects for a given set of cosmological parameters.
By performing many simulations for a whole range of cosmologies and comparing
their results with observations one can constrain these cosmological 
parameters. In practice, in order to break degeneracies between various
parameters one combines weak lensing surveys with other cosmological probes 
(CMBR observations, cluster statistics, etc.) to obtain best constraints.
Note that such a procedure assumes that the actual universe belongs to the 
class of models described by these cosmological parameters. In particular,
one usually assumes a ``cold-dark matter'' (CDM) scenario (most of the matter 
is in the form of an unknown cold and collisionless component, which is 
necessary to explain many astronomical observations like the rotation curves 
of spiral galaxies) where non-linear objects (like galaxies) and large scale
structures (filaments, voids,etc.) build up through the amplification by
gravitational instability of small initial Gaussian density fluctuations, 
while the whole universe follows a ``big-bang'' dynamics described by the
Friedmann-Robertson-Walker metric (Peebles 1993). The agreement of
many observational tests with the predictions derived from such scenarios
strongly suggests this global picture is indeed correct (until we find
a discrepancy which cannot be reconciled) and observational measurements
can be further used to obtain the values of cosmological parameters like
the mean matter density in the universe.

\section{Weak lensing statistics}

\subsection{Power spectrum}

From eq.(\ref{shearmatrix}) we can see that the means $\lag\kappa\rag$ and
$\lag{\bf \gamma}\rag$ vanish (since $\lag\delta\rag=0$ whence 
$\lag\phi\rag=0$). Therefore, the lowest-order statistics we can measure
are second-order moments like $\lag\kappa^2\rag$. In particular, the power 
spectrum $P_{\kappa}(s)$ of the weak lensing convergence, defined in 
two-dimensional Fourier space by $\lag \kappa({\bf s}) \kappa({\bf s'})^*\rag =
\delta_D({\bf s}-{\bf s'}) P_{\kappa}(s)$, is related to the power spectrum
$P_{3D}$ of the three-dimensional density contrast $\delta$, by (Kaiser 1998):
\begin{equation}
P_{\kappa}(s) = \frac{9\Omega_m^2}{4} \frac{H_0^4}{c^4} 
\int_0^{\infty} \frac{{\rm d}\chi}{a^2} \; P_{3D} \left(\frac{s}{{\cal D}};
z\right) \left[ \int_z^{\infty} {\rm d}z_s \; n(z_s) 
\frac{{\cal D}(\chi_s-\chi)}{{\cal D}(\chi_s)} \right]^2 .
\label{Power}
\end{equation}
Here we used the small-angle approximation.
The notations are the same as in eq.(\ref{shearmatrix}), the redshift $z$ 
corresponds to the radial distance $\chi$ while $n(z_s)$ is the galaxy redshift
distribution normalized to unity. We noted $c$ the speed of light, 
$H_0$ the present value of the
Hubble constant and $\Omega_m$ the matter density parameter. From 
eq.(\ref{shearmatrix}) one can check that the power spectrum of the shear 
${\bf \gamma}$ is equal to $P_{\kappa}(s)$ (thus $\lag\kappa^2\rag=
\lag|{\bf\gamma}|^2\rag$). In practice, one does not measure the power 
spectrum $P_{\kappa}(s)$ itself but the variance of the shear smoothed with
a suitable filter. Thus, one can consider the smoothed convergence, shear
components and the aperture mass $\Map$ (Schneider et al. 1998a) defined by:
\begin{equation}
\kappa_s= \int{\rm d}^2\theta \; U({\bf \theta}) \kappa({\bf \theta}) , \;\;
\gamma_{is}= \int{\rm d}^2\theta \; U({\bf \theta}) \gamma_i({\bf \theta}) , 
\;\; \Map= \int{\rm d}^2\theta \; Q({\bf \theta}) \gamma_t({\bf \theta}) ,
\label{filters}
\end{equation}
where $U({\bf \theta})$ is a top-hat of size $\theta_s$ while $Q({\bf \theta})$
is a compensated filter (of zero mean) and $\gamma_t$ is the tangential shear
(with respect to the window center). The advantage of the compensated quantity
$\Map$ is that it provides a pass-band filter which directly probes the
convergence power at wavenumber $s \sim 1/\theta_s$. Indeed, since the filter
$Q({\bf \theta})$ has a zero mean 
(i.e. $\int{\rm d}^2\theta \; Q({\bf \theta})=0$) the contribution from 
wavelengths much larger than $\theta_s$ is suppressed, which is not the
case for the observables $\kappa_s$ and $\gamma_{is}$ defined with the top-hat
filter $U$. The drawback is that the variance $\lag\Map^2\rag$ is smaller than
$\lag\kappa_s^2\rag=2\lag\gamma_{is}^2\rag$. Another property of the aperture
mass $\Map$ is that it can be expressed both in terms of the shear, as in
eq.(\ref{filters}), or in terms of the convergence $\kappa$ (Schneider 1996). 
Of course, the variance of all these observables can 
be written in terms of the power spectrum $P_{\kappa}(s)$.

\begin{figure}

\centering
\includegraphics[width=9cm]{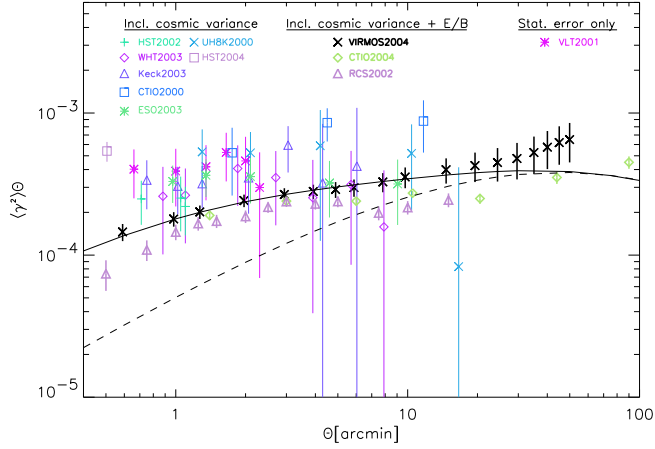}

\caption{A compilation of results from various surveys. The figure shows 
$\lag|\gamma|^2\rag$ times the radius of the smoothing scale $\theta$ as a 
function of smoothing angular scale. The solid line is the best fit model, 
and the dashed line shows the linear prediction 
($\Omega_m=0.3$, $\Omega_{\Lambda}=0.7$, $\sigma_8=0.83$ and source redshift 
$z_s=0.9$). The three thick symbols show the observations where an 
Electric/Magnetic mode separation was carried out, the thin symbols show 
measurements without such separation. It is interesting to see that all 
measurements with no separation  predict a larger $\sigma_8$. The RCS and 
CTIO2004 survey results are from low redshift ($z_s = 0.6$) shallow surveys. 
These measurements were rescaled in order to have an equivalent source 
redshift of $z_s =0.9$ (to match others). 
Figure courtesy Ludovic Van Waerbeke.}
\label{Figsurveys}
\end{figure}

Then, one could divide the survey area on the sky into numerous patches of 
radius $\theta_s$ and perform the average of $\Map^2$ (for instance) over 
all cells. However, this method is not well suited to real weak lensing 
surveys which have a complex geometry with many holes of various sizes, as one
needs to remove the areas contaminated by bright stars or observational 
defects. This makes it impracticable to draw a circular filter across the data.
Hence, one usually defines shear two-point correlations, such as 
$\xi_t=\lag\gamma_t\gamma_t\rag$ and $\xi_r=\lag\gamma_r\gamma_r\rag$
from the tangential and radial components $\gamma_t$ and $\gamma_r$ of the 
shear (the cross-correlation $\lag\gamma_t\gamma_r\rag$ vanishes by 
symmetry), see Schneider, van Waerbeke \& Mellier (2002).
These two-point correlations are measured by summing the shear products
over all galaxy pairs separated by some angle ${\bf\theta}$. This bypasses the
problem raised by the numerous holes present in the survey. Next, the variance
of the shear or of the aperture mass can be expressed as integrals over
$\xi_t$ and $\xi_r$ (Crittenden et al. 2002; 
Schneider, van Waerbeke \& Mellier 2002). 
Alternatively, one can study the shear two-point correlations themselves. 

As in electromagnetism, the tensor field ${\bf\gamma}$ can be decomposed
into an electric ($E$ mode, i.e. gradient) component and a magnetic 
($B$ mode, i.e. curl) component. However, since weak lensing effects are 
produced by a scalar field (the gravitational potential $\phi$) the $B$ mode
is zero. Therefore, any detection of the $B$ mode in the data signals the
presence of systematics or other effects not related to weak lensing like
intrinsic galaxy alignements. It can be shown that the aperture mass $\Map$
given in eq.(\ref{filters}) only selects the $E$ mode, while the same quantity
measured after all galaxies have been rotated by 45 degrees (noted for instance
$M_{\times}$) selects the $B$ mode (Crittenden et al. 2002; 
Schneider, van Waerbeke \& Mellier 2002). Therefore, by computing the variances
$\lag\Map^2\rag$ and $\lag M_{\times}^2\rag$ one can estimate the amount of 
contamination to the weak lensing signal.
Although present surveys find a non-zero $B$ mode it
is sufficiently smaller than the $E$ mode to support the detection of a
real weak lensing signal due to large scale structures. Unfortunately,
it is not clear how the $B$ mode can be substracted from the $E$ mode
to improve the accuracy of the measurements. 
This remains an important issue for weak lensing surveys.

By varying 
the smoothing angle $\theta_s$ or the pair separation $\theta$ one can 
constrain the shape of the convergence power spectrum $P_{\kappa}(s)$ and 
of the underlying matter power spectrum $P_{3D}(k)$ over the range probed 
by the survey, as measured by the usual parameter $\Gamma$. On the other hand, 
as can be seen in eq.(\ref{Power}) the amplitude of the weak lensing signal 
measures a combination of the matter density parameter $\Omega_m$ and of the 
normalization of the matter power spectrum $P_{3D}$ (usually noted 
$\sigma_8$). The dependence on other cosmological parameters is much smaller 
but one must pay attention to the rather strong dependence on the redshift 
of the sources.

After early failed searches for weak lensing signals which  include studies 
mainly using photometric plates  (Kristian 1967; Valdes, Jarvis \& Tyson 1983),
a first attempt with CCD (Mould et al. 1994) only derived an upper limit. 
However Villumsen (1995) using the same data reported a detection, and a few 
years later Schneider et al. (1998b) reported detection of weak lensing. 
Nevertheless, the sky coverage of these studies was probably too small. 
After the initial discovery within a short period of time four groups 
independently confirmed detection of cosmic shear in random patches of the 
sky (Bacon, Refregier \& Ellis 2000; Kaiser, Wilson \& Luppino 2000; 
van Waerbeke et al. 2000; Witman et al. 2000). It is interesting to note that 
all of these initial observations were done using 4m-class telescope. Since 
then there has been an explosion of shear measurements coming from various
observational teams from ground based observations (Bacon et al. 2003; 
Brown et al. 2003; Hamana et al. 2003; Hoekstra et al. 2002; 
Hoekstra, Yee \& Gladders 2002; Jarvis et al. 2002; Maoli et al. 2001;
van Waerbeke et al. 2001a; van Waerbeke et al. 2002) as well as from space 
using the Hubble Space Telescope (see e.g. H\"{a}mmerle
et al. 2002; Rhodes, Refregier \& Groth 2001). Figure \ref{Figsurveys} shows
a compilation of present results from various surveys. Although there is a
clear detection and the various experiments agree with the theoretical
expectations, there is still a some dispersion between different
surveys. In particular, it appears that the separation between $E$ modes and
$B$ modes needs to be considered with care.

A more ambitious goal than measuring a few cosmological parameters like
$(\Omega_m,\sigma_8,\Gamma)$ is to use weak lensing surveys to reconstruct
the matter power spectrum $P_{3D}$ itself, by inverting eq.(\ref{Power}).
This is not an easy task because the integration along the line of sight
implies that a whole range of physical three dimensional wavenumbers $k$
contribute to a given two dimensional wavenumber $s$.
Another problem is that the evolution with redshift $z$ of the matter power 
spectrum $P_{3D}(k;z)$ in eq.(\ref{Power}) only factorizes as 
$P_{3D}(k;z)= D_+(z) P_L(k)$ in the linear regime (which in principles allows 
one to recover the function of one variable $P_L(k)$ from $P_{\kappa}(s)$).
In the non-linear regime this factorization is no longer valid and one needs
to work with the function of two variables $P_{3D}(k;z)$. In practice one
may circumvent this problem by using a mapping from the linear prediction
$D_+(z) P_L(k)$ onto the non-linear power spectrum (or by neglecting the
departures from linear growth) but this procedure introduces some modeling
associated with this mapping (or some approximation). Alternatively, one may
restrict the reconstruction to very large linear scales. An estimate of
the two dimensional converge power spectrum was obtained in Pen et al. (2002) 
and Brown et al. (2003) while Pen et al. (2003a) derived the three dimensional 
matter density power spectrum.

\subsection{Non-Gaussianities}

To break the degeneracy between the parameters $\Omega_m$ and 
$\sigma_8$ present in two-point statistics one can combine weak lensing 
observations with other cosmological probes
(like CMBR) or consider higher order moments of weak lensing observables.
Indeed, even if the initial conditions are Gaussian, since the dynamics is
non-linear (this is an unstable self-gravitating expanding system) 
non-Gaussianities develop and in the non-linear regime the density field 
becomes strongly non-Gaussian. This can be seen from the constraints
$\lag\delta\rag=0$ and $\delta \geq -1$ (because the matter density $\rho$ 
is positive) which implie that in the highly non-linear regime 
($\lag\delta^2\rag\gg 1$) the probability distribution of the density contrast 
$\delta$ must be far from Gaussian. In the quasi-linear regime where
a perturbative approach is valid (with Gaussian initial conditions) one 
can show that the skewness $S_3=\lag\kappa_s^3\rag/\lag\kappa_s^2\rag^2$ 
is independent of the density power spectrum normalization $\sigma_8$
(the same property is valid for other observables like $\Map$ which are linear
over the matter density field). Therefore, by measuring the second and third 
order moments of the convergence or of the aperture mass at large angular 
scales one can obtain a constraint on $\Omega_m$
(Bernardeau, van Waerbeke \& Mellier 1997).
Alternatively, from the skewness of weak lensing observables one
can derive the skewness of the matter density field in the linear regime
and check that the scenario of the growth of large scale structures through
gravitational instability from initial Gaussian conditions is valid.

Again, because of the numerous holes within the survey area one first computes 
shear three-point correlations (called the bispectrum in Fourier space) 
by summing over galaxy triplets and next 
writes $\lag\Map^3\rag$ as an integral over these three-point correlations.
Applying this method 
to the Virmos-Descart data Pen et al. (2003b) were able to detect
$S_3$ and to infer an upper bound $\Omega_m<0.5$ by comparison with
simulations. Alternatively, one can study the shear three-point correlation 
itself, averaged over some angular scale. This can be optimised by identifying
typical shear patterns (such as the flow of the shear vector ${\bf\gamma}$
around a galaxy pair) in order to select the three point product which yields
the larger signal (recall that $\gamma$ has two components which offers many
choices). Bernardeau et al. (2002) obtained in this fashion the first 
detection of non-Gaussianity in a weak lensing survey.
Next, one could measure higher order moments of weak lensing
observables. Note that for the shear components odd order moments vanish by
symmetry so that one needs to consider the fourth-order moment to go beyond 
the variance (Takada \& Jain 2002). However, higher order moments are 
increasingly noisy (Valageas, Munshi \& Barber 2005) so that it has not been 
possible to go beyond the skewness yet.

In practice, most of the angular range probed by weak lensing surveys is
actually in the transition domain from the linear to highly non-linear regimes
(from $10'$ down to $1'$). Therefore, it is important to have a reliable
prediction for these mildly and highly non-linear scales, once the cosmology
and the initial conditions are specified. Since there is no rigorous 
analytical framework to investigate this regime numerical simulations play
a key role to obtain the non-linear evolution of the matter power spectrum
and of higher order statistics (Peacock \& Dodds 1996; Smith et al. 2003). 
Based on these simulation results it is
possible to build analytical models which can describe the low order moments
of weak lensing observables or their full probability distribution. This can
be done through a hierarchical {\it ansatz} where all higher-order density 
correlations are expressed in terms of the two-point correlation 
(Barber, Munshi \& Valageas 2004; Valageas, Munshi \& Barber 2005).
Then, the probability distribution of weak lensing observables can be directly
written in terms of the probability distribution of the matter density. 
In some cases the mere existence of this relationship allows one to 
discriminate between analytical models for the density field which are very 
similar (Munshi, Valageas \& Barber 2004).
Alternatively, one can use a halo model where the matter distribution
is described as a collection of halos (Cooray \& Sheth 2002) and the low 
order moments of weak lensing observables can be derived by averaging over 
the statistics of these halos (Takada \& Jain 2003a,b). On the other hand,
one can use weak lensing to constrain halo properties and to detect 
substructures (Dolney, Jain \& Takada 2004).

\subsection{Weak lensing probe of astrophysics at small scales}

Since weak lensing surveys directly probe the matter distribution in the
universe, they can be used together with galaxy surveys to estimate
the bias of galaxies and the matter-galaxy cross-correlation.
Present generation surveys are already proving useful in this direction 
(Simon et al. 2004). Future weak lensing surveys will have a direct impact on 
galaxy formation scenarios by probing the bias associated with various 
galaxy types and its evolution with redshift $z_s$.

\section{Problems to overcome}

Weak lensing surveys are only beginning to address cosmological questions
like the value of cosmological parameters. In order to exploit gravitational
lensing effects for cosmological purposes with a good accurvay we still need 
to improve our control of various sources of errors. We list below a few of
these issues.

{\it Intrinsic galaxy alignment.} Weak lensing measurements can be 
contaminated by a possible intrinsic alignment of galaxy ellipticites, which
could be induced by tidal fields (see Heavens 2001). Theoretical estimates of 
such an effect have been performed using numerical (Jing 2002; 
Croft \& Metzler 2000; Heavens, Refregier \& Heymans 2000) and analytical 
techniques (Catalan, Kamionkowski \& Blandford 20001; 
Crittenden et al. 2001, 2002; Lee \& Pen 2001; 
Mackey, White \& Kamionkowski 2002). Measurements of intrinsic ellipticity 
correlation of galaxies have also been attempted (Pen, Lee \& Seljak 2000; 
Brown et al. 2002). At present there is no clear agreement among various 
analytical predictions regarding the amplitude of this effect, although one
can safely expect that shallower surveys should be more affected than 
deeper surveys. To suppress this possible source of errors in future surveys
one will use photometric redshifts to select galaxies which are separated 
in redshift space. However, in a recent study Hirata \& Seljak (2004) 
have shown that even the galaxy intrinsic ellipticity and the shear field 
can be correlated which would make things further complicated.

{\it Redshift distribution.} Weak lensing effects depend on the source
redshift which needs to be known with a good accuracy in order to obtain
tight constraints on cosmology. This is a difficult task because of the
depth of weak lensing surveys. Moreover, any dependence of the source
redshift distribution on the direction on the sky could affect the comparison
of observations with theory. Over the course of next few years, photometric 
redshift surveys (e.g. COMBO17 \footnote{http://www.mpia-hd.mpg.de/COMBO/}) 
are going to be increasingly prevalent. On the other hand, if the survey has
a broad source redshift distribution, background sources can
be lensed by foreground galaxies or by matter density peaks correlated with
foreground sources. This effect yields additional terms to the weak lensing 
signal which must be taken into account (Bernardeau 1998a; Hamana et al. 2002).

{\it PSF correction.} The most serious issue (for the observational part)
is the distortion induced by the PSF. At present the KSB method allows an
accuracy of a few percent. In order to reach higher precision 
one needs to develop more efficient techniques. Space telescopes
avoid the atmospheric turbulence and show a more stable PSF but they still
require a significant correction.

{\it Non-linear evolution.}  Most of the angular scales measured by 
weak lensing surveys ($<10'$) probe the non-linear regime of gravitational
clustering. Since no analytical methods are available,
one needs to use numerical simulations to obtain theoretical predictions
which can be compared with weak lensing observations 
(van Waerbeke et al. 2001b). Current simulations
only provide an accuracy of the order of $10\%$ for two-point statistics
while higher order statistics are increasingly difficult to obtain. 
Recent works such as Smith et al. (2003) already show an improvement over
previous estimates but the development of more powerful simulations remains
necessary.

{\it Numerical simulations.} Simulation techniques also need to
be generalised and improved upon to quantify the corrections due to many 
realistic details such as source-lens coupling, source clustering and redshift
distribution. Contributions from the finite size of the surveys and intrinsic
noise can be investigated in great details to optimise returns from future 
surveys. The construction of such mock catalogues will pave the way for 
analytical modelling of various realistic 
systematics\footnote{For more discussions on simulations including 
many nice plots see http://www.cfht.hawaii.edu/News/Lensing}.

\section{Prospects}

Although CMBR observations together with other probes like SNIa experiments
and cluster statistics have already provided impressive information on
some key cosmological issues, weak lensing surveys are a promising tool for
cosmological purposes. First, used in combination with other observations
like CMBR missions they can help break degeneracies between cosmological
parameters and they can tighten the constraints on models. Second, they
offer a unique method to probe the matter distribution on quasi-linear to
non-linear scales which cannot be done by other techniques. We list below
a few of the goals of future surveys.

{\it Power spectrum reconstruction.} Using the redshift position of
the source together with the weak lensing signal itself one can study
the evolution with redshift of the matter distribution. It may even be 
possible to map the three dimensional matter distribution 
(Hu \& Keton 2002; Massey et al. 2004; Taylor 2001). 

{\it Cosmological parameters.} The measured cosmic signals from 
small weak lensing surveys can be used to provide constraints
on the cosmological parameters $\Omega_m, \sigma_8$ and $\Gamma$ 
(the source redshift distribution also plays a very 
important role). A much bigger survey with much higher sky coverage allows
to constrain many more cosmological 
parameters (Hu \& Tegmark 1999), see also Tereno et al. (2005)
for a recent joint analysis of weak lensing surveys with current 
WMAP-1 year and CBI data employing Monte Carlo Markov Chain calculation of a 
seven parameter model, which is more realistic than previously employed 
Fisher matrix based analysis. A principal component analysis which probes 
various parameter degeneracies is also presented. Next generation surveys 
will be independently able to probe such issues without having to assume a 
specific structure formation scenarios with a primordial power law spectrum.
They will also constrain the evolution of the dark energy and
test quintessence models (Benabed \& van Waerbeke 2004; Hu \& Jain 2004).
In particular, lensing tomography (i.e. binning galaxy sources in redshift
space) can prove useful to tighten the constraints on cosmology (Hu 1999;
Takada \& Jain 2004).

{\it Primordial non-Gaussianities.} Results from future surveys will also 
be very useful in constraining primordial non-Gaussianity predicted 
by some early universe theories. Generalised theories of gravity can 
have very different predictions regarding gravity induced non-Gaussianites 
as compared to GR, which can also be probed using future data. A joint 
analysis of power spectrum and bi-spectrum from weak lensing 
surveys thus will provide a very powerful way to constrain, not only 
cosmological parameters, but early universe theories and alternative 
theories of gravitation (see also Schmid, Uzan \& Riazuelo 2005). 
Thus, in a recent work Bernardeau (2004) has studied 
the possibility of constraining higher-dimensional gravity
from cosmic shear three-point correlation function.

{\it Future surveys.} The constraints on cosmology brought by weak lensing 
measurements will gradually improve as ongoing surveys are fully exploited,
such as the CFHT (van Waerbeke, Mellier \& Hoekstra 2005), the Deep Lens 
Survey (Wittmann et al. 2002), the SDSS (Sheldon et al. 2004). Accurate
 measures of cosmological parameters and tests of various cosmological 
scenarios will be possible with future missions, such as the CFHT Legacy 
Surveys (Tereno et al. 2005),
the Visible and Infrared Survey Telescope for Astronomy or VISTA 
(Taylor et al. 2003), the Large aperture Synoptic Survey Telescope or LSST 
(Tyson et al. 2002a,b), novel Panoramic Survey Telescope and Rapid Response 
System or Pan-STARRS (Kaiser, Tonry \& Luppino 2000), Supernova Acceleration 
Probe satellite or SNAP (Perlmutter et al. 2003) and Advanced Camera for 
Surveys (ACS) on HST. Ground based and space-based observations have a 
complimentary role to play in near future. While space based observations 
can provide a stable PSF and hence a reduced level of systematics ground based 
observations can survey a larger fraction of the sky. However, a high level 
of degeneracy among various lensing observables implies that only a small 
number of independent linear combinations of them can be extracted from future 
surveys. On the other hand, even smaller surveys will be very helpful
in improving cosmological constraints when analysed jointly with external 
data-sets such as data from all-sky cosmic microwave background surveys 
(Hu \& Tegmark 1999).

Weak lensing of background galaxy samples is limited only to a low source 
redshift($z_s \approx 2$). However as pointed out by various authors 
(e.g. Bernardeau 1997, 1998b; Cooray \& Kesden 2003;
Hirata \& Seljak 2003; Seljak 1996; van Waerbeke, Bernardeau \& Benabed 2000;
Zaldarriaga \& Seljak 1999) 
weak lensing of CMBR can be used to study the matter clustering
all the way up to recombination era. It was also realised thanks to these 
studies that distinct non-Gaussian signal left due to lensing can actually 
be used to reconstruct the foreground mass distribution. With forthcoming 
high resolution CMB missions such as Planck Surveyor such a program will be 
made feasible.

Possibilities of weak lensing studies using  radio surveys such as FIRST have 
also been studied (Kamionkowski et al. 1997; Refregier et al. 1998). Upcoming 
radio surveys such as Low Frequency Array or LOFAR and Square Kilometer 
Array (SKA) will provide unique opportunity in this direction 
(see e.g. Schneider 1999).

\section{Conclusions}

Next generation of cosmic shear surveys with MEGACAM at CFHT or VISTA at 
Paranal or even space based panoramic cameras will improve by order of 
magnitude in detail and precision. This will eventually lead to projected 
mass reconstruction, similar to APM galaxy surveys. By allowing measurement 
of higher order correlation functions thanks to huge sky coverage and low shot 
noise, it will break the degeneracy between cosmological parameters such as 
$\sigma_8$ and $\Omega_m$. Using priors from external data set such as recent 
all sky CMB experiments, weak lensing experiments can already put strong 
constraints on extended set of cosmological parameters such as the shape 
parameter $\Gamma$ of power spectrum or the primordial spectral slope 
$n_s$ and running of the spectral index. As surveys get bigger probing 
larger angular scales will be easier and as these scales are free from 
non-linearites they will be very useful to probe the background cosmological 
dynamics. With photometric redshifts it will also be possible to make 3D 
dark matter maps. On a longer timescales very large surveys will start 
probing scales larger than $10$ degrees which will eventually permit us to 
constraint $\Omega_{\Lambda}$ or any quintessence fields. 
While at large angular scale weak lensing measurements will increasingly 
focus on cosmological dynamics and nature of dark energy, small angular scale 
measurements will give us clues to the clustering of baryons relative to 
dark matter distributions. This will be possible by comparing weak lensing 
maps against galaxy surveys.

\begin{acknowledgements}

\noindent Reference: DM would like to thank Adam Amara, Andrew Barber, 
Alan Heavens, Yun Wang, Lindsay King, Martin Kilbinger, Patrick Simon and 
George Efstathiou for useful discussions. We would like to thank Patrick 
Simon, Stephane Colombi and Ludovic Van Waerbeke to make copies of their 
plots and figures available for this review. It is a pleasure for DM to thank 
members of Cambridge Planck Analysis Center. DM was funded by PPARC grant RG28936.

\end{acknowledgements}

\end{document}